\documentclass[a4paper,12pt]{article}
\usepackage{amssymb,graphicx}

\newtheorem{theorem}{Theorem}

\begin{document}

\author{
Dumitru B\u{a}leanu\\
\small{\c{C}ankaya University,}\\
\small{Department of Mathematics {\&} Computer Science,}\\
\small{\"{O}gretmenler Cad. 14 06530, Balgat -- Ankara, Turkey,}\\
\small{{\&} Institute of Space Sciences, M\u{a}gurele -- Bucure\c{s}ti, Romania}\\
\small{e-mail address: dumitru@cankaya.edu.tr}\\
Octavian G. Mustafa\\
\small{University of Craiova, }\\
\small{Faculty of Exact Sciences,}\\
\small{A{.}~I{.} Cuza 13, 200534 Craiova, Romania}\\
\small{e-mail address: octaviangenghiz@yahoo.com}\\
and\\
Donal O'Regan\\
\small{National University of Ireland,}\\
\small{School of Mathematics, Statistics and Applied Mathematics,}\\
\small{Galway, Ireland}\\
\small{e-mail address: donal.oregan@nuigalway.ie}
}

\title{On a fractional differential equation\\
 with infinitely many solutions}
\date{}
\maketitle

\noindent{\bf Abstract} We present a set of restrictions on the fractional differential equation $x^{(\alpha)}(t)=g(x(t))$, $t\geq0$, where $\alpha\in(0,1)$ and $g(0)=0$, that leads to the existence of an infinity of solutions starting from $x(0)=0$. The operator $x^{(\alpha)}$ is the Caputo differential operator.

\section{Introduction}
The issue of multiplicity for solutions of an initial value problem that is associated to some nonlinear differential equation is essential in the modeling of complex phenomena. 

Typically, when the nonlinearity of an equation is not of Lipschitz type \cite{agarw_carte}, there are only a few techniques to help us decide whether an initial value problem has more than one solution. As an example, the equation $x^{\prime}=f(x)=\sqrt{x}\cdot\chi_{(0,+\infty)}(x)$ has an infinity of solutions $x_{T}(t)=\frac{(t-T)^2}{4}\cdot\chi_{(T,+\infty)}(t)$ defined on the nonnegative half-line which start from $x(0)=0$. Here, by $\chi$ we denote the characteristic function of a Lebesgue-measurable set.

An interesting classical result \cite{kamke,agarw_carte}, which generalizes the example, asserts that the initial value problem
\begin{eqnarray}
\left\{
\begin{array}{ll}
x^{\prime}(t)=g(x(t)),&t\geq0,\\
x(0)=x_{0},&x_{0}\in\mathbb{R},
\end{array}
\right.\label{clasic_01}
\end{eqnarray}
where the continuous function $g:\mathbb{R}\rightarrow\mathbb{R}$ has a zero at $x_0$ and is positive everywhere else, possesses an infinity of solutions if and only if $\int_{x_{0}+}^{}\frac{du}{g(u)}<+\infty$.

Recently, variants of this result have been employed in establishing various facts regarding some mathematical models \cite{must2005,must2009}. In particular, if the function $g$ is allowed to have two zeros $x_{0}<x_{1}$ while remaining positive everywhere else and
\begin{eqnarray*}
\int_{x_{0}+}^{}\frac{du}{g(u)}<+\infty,\quad\int^{x_{1}-}_{}\frac{du}{g(u)}=+\infty,
\end{eqnarray*}
then the problem (\ref{clasic_01}) has an infinity of solutions $(x_{T})_{T>0}$ such that $\lim\limits_{t\rightarrow+\infty}x_{T}(t)\allowbreak=x_{1}$.

Our intention in the following is to discuss a particular case of the above non-uniqueness theorem in the framework of fractional differential equations. To the best of our knowledge, the result has not been established in its full generality yet for any generalized differential equation. We mention at this point the closely connected investigation \cite{bmr_2012_1}.

In the last number of years, it became evident that differential equations of non-integer order, also called \textit{fractionals} (FDE's), can capture better in models many of the relevant features of complex phenomena from engineering, physics or chemistry, see the references in \cite{baleanuetal,baleanu1,baleanu2,baleanu3,miller,podlubny,trujillo,bmr2012,bmr2012a}. 

Let us consider a function $h\in C^{1}(I,\mathbb{R})\cap C(\overline{I},\mathbb{R})$ with $\lim\limits_{t\searrow0}[t^{1-\alpha}h^{\prime}(t)]\in\mathbb{R}$ for some $\alpha\in(0,1)$, where $I=(0,+\infty)$. The \textit{Caputo derivative} of order $\alpha$ of $h$ is defined as
\begin{eqnarray*}
h^{(\alpha)}(t)=\frac{1}{\Gamma(1-\alpha)}\cdot\int_{0}^{t}\frac{h^{\prime}(s)}{(t-s)^{\alpha}}ds,\quad t\in I,
\end{eqnarray*}
where $\Gamma$ is Euler's function Gamma, cf{.} \cite[p{.} 79]{podlubny}. To have an idea about the eventual smoothness of this quantity, we mention that by letting the function $h^{\prime}$ be at least absolutely continuous \cite[Chapter 7]{rudin} the (usual) derivative of $h^{(\alpha)}$ will exist almost everywhere with respect to the Lebesgue measure $m$ on $\mathbb{R}$, see \cite[p{.} 35, Lemma 2{.}2]{samko}. Further, we have that
\begin{eqnarray}
h(t)=h(0)+\frac{1}{\Gamma(\alpha)}\int_{0}^{t}\frac{h^{(\alpha)}(s)}{(t-s)^{1-\alpha}}ds,\quad t\in I,\label{integro}
\end{eqnarray}
provided that $h^{(\alpha)}$ is in $L^{\infty}(m)$.

The initial value problem we investigate in this paper is
\begin{eqnarray}
\left\{
\begin{array}{ll}
x^{(\alpha)}(t)=g(x(t)),&t\in I,\\
x(0)=0,&{}
\end{array}
\right.\label{iniprob}
\end{eqnarray}
where the function $g:\mathbb{R}\rightarrow\mathbb{R}$ is continuous, $g(0)=0$ and $g(u)>0$ when $u\in(0,1]$. Further restrictions will be imposed on $g$ to ensure that $\int_{0+}^{}\frac{du}{g(u)}<+\infty$.

By means of (\ref{integro}), we deduce that
\begin{eqnarray*}
x(t)&=&\int_{0}^{t}x^{\prime}(s)ds\\
&=&\frac{1}{\Gamma(\alpha)}\int_{0}^{t}\frac{1}{(t-s)^{1-\alpha}}\left[\frac{1}{\Gamma(1-\alpha)}\int_{0}^{s}\frac{x^{\prime}(\tau)}{(s-\tau)^{\alpha}}d\tau\right]ds
\end{eqnarray*}
and so the problem (\ref{iniprob}) can be recast as
\begin{eqnarray}
\left\{
\begin{array}{ll}
y(t)=g\left(\int_{0}^{t}\frac{y(s)}{(t-s)^{\beta}}ds\right),&t\geq0,\\
y(0)=0,&{}
\end{array}
\right.\label{ce_rezolv}
\end{eqnarray}
where $y=x^{(\alpha)}$, $\beta=1-\alpha$ and the (general) function $g$ has absorbed the constant $\frac{1}{\Gamma(\alpha)}$. 

In the next section, we look for a family $(y_{T})_{T>0}$, with $y_{T}\in C([0,1],\mathbb{R})$, of (non-trivial) solutions to (\ref{ce_rezolv}).

\section{Infinitely many solutions to (\ref{ce_rezolv})}

We start by noticing that the function $f:[0,1]\rightarrow(0,1)$ with the formula $f(x)=\frac{1+x}{2+x-\beta}=1-\frac{1-\beta}{2+x-\beta}$ is increasing. Introduce now the numbers $\delta_{1}$ and $\delta_{2}$ such that
\begin{eqnarray}
f(1)=\frac{2}{3-\beta}>\delta_{1}\geq\delta_{2}>f(0)=\frac{1}{2-\beta}.\label{aux_4}
\end{eqnarray}
Obviously, $\delta_{1},\,\delta_{2}\in(0,1)$.

Set $c_{1}\leq c_{2}$ in $(0,+\infty)$ and assume that
\begin{eqnarray}
c_{1}\cdot y^{\delta_{1}}\leq g(y)\leq c_{2}\cdot y^{\delta_{2}},\quad y\in[0,1].\label{restr_1_g}
\end{eqnarray}
As a by-product, $\int_{0+}^{1}\frac{dy}{g(y)}\leq\frac{c_{1}^{-1}}{1-\delta_{1}}<+\infty$. Further, suppose that there exists $c>0$ such that
\begin{eqnarray}
\vert g(y_{2})-g(y_{1})\vert\leq\frac{c}{\left(\min\{y_{1},y_{2}\}\right)^{1-\delta_{1}}}\cdot\vert y_{2}-y_{1}\vert,\quad y_{1},\,y_{2}\in(0,1].\label{restr_2_g}
\end{eqnarray}
The latter condition has been inspired by the analysis in \cite{must1}.

Introduce now the numbers $Y_{1},\,Y_{2}\geq1$ and $T\in(0,1)$ with
\begin{eqnarray}
(Y_{1}+Y_{2})(1-T)^{2-\beta}<1-\beta\label{domeniul_lui_y_1}
\end{eqnarray}
and
\begin{eqnarray}
8Y_{1}\leq c_{1}\leq c_{2}\leq(1-\beta)Y_{2}^{1-\delta_{2}}\label{domeniul_lui_y_2}
\end{eqnarray}
and
\begin{eqnarray}
k=\frac{c}{(1-\beta)^{\delta_{1}}}\cdot\left(\frac{8}{Y_{1}}\right)^{1-\delta_{1}}<1.\label{domeniul_lui_y_2_1}
\end{eqnarray}
These will be used in the following for describing the solution $y_{T}$.

Several simple estimates, of much help in the proof of our result, are established next. Notice first that, via the change of variables $s=T+u(t-T)$, we get
\begin{eqnarray}
\int_{T}^{t}\frac{(s-T)^{1+\varepsilon}}{(t-s)^{\beta}}&=&(t-T)^{2+\varepsilon-\beta}\cdot\int_{0}^{1}\frac{u^{1+\varepsilon}}{(1-u)^{\beta}}du\nonumber\\
&=&B(2+\varepsilon,1-\beta)(t-T)^{2+\varepsilon-\beta},\label{aux_1}
\end{eqnarray}
where $\varepsilon\in(0,1)$ and $B$ represents Euler's function Beta \cite{podlubny}. Also,
\begin{eqnarray}
B(2+\varepsilon,1-\beta)&\geq&\int_{\frac{1}{2}}^{1}\frac{u^{1+\varepsilon}}{(1-u)^{\beta}}du\geq\frac{1}{2^{1+\varepsilon}}\cdot\int_{\frac{1}{2}}^{1}\frac{du}{(1-u)^{\beta}}\nonumber\\
&=&\frac{2^{-(2+\varepsilon-\beta)}}{1-\beta}\geq\frac{1}{8(1-\beta)}\label{aux_2}
\end{eqnarray}
and
\begin{eqnarray}
B(2+\varepsilon,1-\beta)\leq\int_{0}^{1}\frac{du}{(1-u)^{\beta}}=\frac{1}{1-\beta}.\label{aux_3}
\end{eqnarray}

Now, returning to (\ref{aux_4}), there exist $\varepsilon_{1},\,\varepsilon_{2}\in(0,1)$, with $\varepsilon_{1}\geq\varepsilon_{2}$, such that
\begin{eqnarray}
f(\varepsilon_{1})=\delta_{1}\geq\delta_{2}=f(\varepsilon_{2}).\label{aux_5}
\end{eqnarray}
In particular, $1-\delta_{1}=\frac{1-\beta}{2+\varepsilon_{1}-\beta}$ and, by means of (\ref{aux_3}),
\begin{eqnarray}
&&Y_{1}B(2+\varepsilon_{1},1-\beta)(1-T)^{2+\varepsilon_{1}-\beta}\nonumber\\
&&+Y_{2}B(2+\varepsilon_{2},1-\beta)(1-T)^{2+\varepsilon_{2}-\beta}<1.\label{domeniul_lui_y_6}
\end{eqnarray}

Taking into account (\ref{aux_2}) and (\ref{domeniul_lui_y_2}), we deduce that
\begin{eqnarray*}
Y_{1}^{1-\delta_{1}}\leq c_{1}\left[\frac{1}{8(1-\beta)}\right]^{\delta_{1}}\leq c_{1}B(2+\varepsilon_{1},1-\beta)^{\delta_{1}},
\end{eqnarray*}
which leads to
\begin{eqnarray}
Y_{1}\leq c_{1}[Y_{1}B(2+\varepsilon_{1},1-\beta)]^{\delta_{1}},\label{domeniul_lui_y_3}
\end{eqnarray}
and, via (\ref{aux_3}),
\begin{eqnarray*}
c_{2}B(2+\varepsilon_{2},1-\beta)^{\delta_{2}}\leq\frac{c_{2}}{(1-\beta)^{\delta_{2}}}\leq\frac{c_{2}}{1-\beta},
\end{eqnarray*}
which implies that
\begin{eqnarray}
c_{2}[Y_{2}B(2+\varepsilon_{2},1-\beta)]^{\delta_{2}}\leq Y_{2}.\label{domeniul_lui_y_4}
\end{eqnarray}

Let the set ${\cal Y}\subset C([T,1],\mathbb{R})$ be given by the double inequality
\begin{eqnarray}
Y_{1}(t-T)^{1+\varepsilon_{1}}\leq y(t)\leq Y_{2}(t-T)^{1+\varepsilon_{2}},\quad t\in[T,1],\,y\in{\cal Y}.\label{domeniul_lui_y_5}
\end{eqnarray}

Observe that, by means of (\ref{domeniul_lui_y_3}), (\ref{aux_1}), (\ref{restr_1_g}),
\begin{eqnarray}
Y_{1}B(2+\varepsilon_{1},1-\beta)(t-T)^{2+\varepsilon_{1}-\beta}\leq\int_{T}^{t}\frac{y(s)}{(t-s)^{\beta}}ds\label{aux_0xxx}
\end{eqnarray}
and
\begin{eqnarray}
Y_{1}(t-T)^{1+\varepsilon_{1}}&\leq&c_{1}[Y_{1}B(2+\varepsilon_{1},1-\beta)]^{\delta_{1}}\cdot(t-T)^{1+\varepsilon_{1}}\nonumber\\
&=&c_{1}\left[Y_{1}B(2+\varepsilon_{1},1-\beta)(t-T)^{2+\varepsilon_{1}-\beta}\right]^{\delta_{1}}\label{aux_0x}\\
&\leq& g\left(\int_{T}^{t}\frac{y(s)}{(t-s)^{\beta}}ds\right).\nonumber
\end{eqnarray}
Similarly, via (\ref{domeniul_lui_y_4}),
\begin{eqnarray}
\int_{T}^{t}\frac{y(s)}{(t-s)^{\beta}}ds\leq Y_{2}B(2+\varepsilon_{2},1-\beta)(t-T)^{2+\varepsilon_{2}-\beta}\label{aux_0xx}
\end{eqnarray}
and
\begin{eqnarray*}
g\left(\int_{T}^{t}\frac{y(s)}{(t-s)^{\beta}}ds\right)&\leq&c_{2}[Y_{2}B(2+\varepsilon_{2},1-\beta)]^{\delta_{2}}\cdot(t-T)^{1+\varepsilon_{2}}\\
&\leq&Y_{2}(t-T)^{1+\varepsilon_{2}}.
\end{eqnarray*}

In conclusion, the mapping $t\mapsto g\left(\int_{T}^{t}\frac{y(s)}{(t-s)^{\beta}}ds\right)$ is a member of ${\cal Y}$ whenever $y\in{\cal Y}$. Also, taking into account (\ref{domeniul_lui_y_6}), we deduce that the quantities $y=Y_{1}B(2+\varepsilon_{1},1-\beta)(t-T)^{2+\varepsilon_{1}-\beta}$ from (\ref{aux_0x}) and $y=Y_{2}B(2+\varepsilon_{2},1-\beta)(t-T)^{2+\varepsilon_{2}-\beta}$ from (\ref{aux_0xx}) belong to $[0,1]$ as imposed in (\ref{restr_1_g}).

We are now ready to state and prove our main result.

\begin{theorem}
Assume that the nonlinearity $g$ of (\ref{ce_rezolv}) satisfies the restrictions (\ref{aux_4}), (\ref{restr_1_g}), (\ref{restr_2_g}). Given the numbers $Y_{1},\,Y_{2},\,T$ subject to (\ref{domeniul_lui_y_1}), (\ref{domeniul_lui_y_2}), (\ref{domeniul_lui_y_2_1}) and the set ${\cal Y}={\cal Y}(Y_{1},Y_{2},T)$ from (\ref{domeniul_lui_y_5}), the problem (\ref{ce_rezolv}) has a unique solution $y_{T}$ in ${\cal Y}$.
\end{theorem}

\textbf{Proof.} The operator ${\cal O}:{\cal Y}\rightarrow{\cal Y}$ with the formula
\begin{eqnarray*}
{\cal O}(y)(t)=g\left(\int_{T}^{t}\frac{y(s)}{(t-s)^{\beta}}ds\right),\quad y\in{\cal Y},\,t\in[T,1],
\end{eqnarray*}
is well defined.

The typical $\sup$--metric $d(y_{1},y_{2})=\sup\limits_{t\in[T,1]}\vert y_{1}(t)-y_{2}(t)\vert$ provides the set ${\cal Y}$ with the structure of a complete metric space.

Taking into account (\ref{restr_2_g}), (\ref{aux_0xxx}) and (\ref{domeniul_lui_y_2_1}), we get
\begin{eqnarray*}
&&\vert{\cal O}(y_{1})(t)-{\cal O}(y_{2})(t)\vert\\
&&\leq\frac{c}{\left[Y_{1}B(2+\varepsilon_{1},1-\beta)(t-T)^{2+\varepsilon_{1}-\beta}\right]^{1-\delta_{1}}}\cdot\int_{T}^{t}\frac{\vert y_{1}(s)-y_{2}(s)\vert}{(t-s)^{\beta}}\\
&&\leq c\left[\frac{8(1-\beta)}{Y_{1}}\right]^{1-\delta_{1}}\cdot\frac{1}{(t-T)^{1-\beta}}\cdot d(y_{1},y_{2})\int_{T}^{t}\frac{ds}{(t-s)^{\beta}}\\
&&=k\cdot d(y_{1},y_{2}),\quad y_{1},\,y_{2}\in{\cal Y}.
\end{eqnarray*}

The operator ${\cal O}$ being thus a contraction, its fixed point $y_{T}$ in ${\cal Y}$ is the solution we are looking for. Notice that $y_{T}$ is identically null in $[0,T]$. $\square$

\textbf{Acknowledgment.} The work of the second author has been supported by a grant of the Romanian National Authority for Scientific Research, CNCS –-- UEFISCDI, project number PN-II-ID-PCE-2011-3-0075.


\begin{thebibliography}{99}

\bibitem{agarw_carte} R.P. Agarwal, V. Lakshmikantham, Uniqueness and nonuniqueness criteria for ordinary differential equations, World Scientific, New Jersey, 1993

\bibitem{baleanuetal}
 D. B\u{a}leanu, K. Diethelm, E. Scalas, J.J. Trujillo, Fractional calculus models and numerical methods. Series on Complexity, Nonlinearity and Chaos,  World Scientific, Boston, 2012

\bibitem{baleanu1} D. B\u{a}leanu, T. Avkar, Lagrangians with linear velocities wi\-thin Rie\-mann-Liouville fractional derivatives, Nuovo Cimento B{.} 119 (2004), 73--79

\bibitem{bmr_2012_1}D. B\u{a}leanu, O.G. Mustafa, D. O'Regan, A uniqueness criterion for fractional differential equations with Caputo derivative, Nonlinear Dyn{.}, in press

\bibitem{baleanu2}
S. Bhalekar, V. Daftardar-Gejji, D. B\u{a}leanu, R.L. Magin, Transient chaos in fractional Bloch equations, Comput. Math. Appl.,\\
\verb+http://dx.doi.org/10.1016/j.camwa.2012.01.069+ (2012)
\bibitem{bmr2012}
H. Delavari, D. B\u{a}leanu, J. Sadati, Stability analysis of Caputo fractional-order nonlinear systems revisited,  Nonlin. Dyn. 67 (2012), 2433--2439 
\bibitem{bmr2012a}
O.P. Agrawal, O. Defterli, D. B\u{a}leanu, Fractional optimal control problems with several state and control variables, J. Vibr. Contr. 16 (2010), 1967--1976

\bibitem{baleanu3}
M.A.E. Herzallah, D. B\u{a}leanu, Fractional Euler-Lagrange equations revisited, Nonlinear Dyn{.},\\
\verb+http://dx.doi.org/10.1007/s11071-011-0319-5+ (2012)

\bibitem{kamke}E. Kamke, Differentialgleichungen, l\"{o}sungsmethoden und l\"{o}sungen I, Akad{.} Verlag{.}, Leipzig, 1961

\bibitem{trujillo}  A.A. Kilbas, H.M. Srivastava, J.J. Trujillo, Theory and applications of
fractional differential equations, North-Holland, New York, 2006

\bibitem{miller} K.S. Miller, B. Ross, An introduction to the fractional calculus and fractional
differential equations, Wiley {\&} Sons, New York, 1993

\bibitem{must2005}O.G. Mustafa, Initial value problem with infinitely many linear-like solutions for a second-order differential equation, Appl{.} Math{.} Lett{.} 18 (2005), 931--934

\bibitem{must2009}O.G. Mustafa, On smooth traveling waves of an integrable two-component Camassa-Holm shallow water system, Wave Motion 46 (2009), 397--402

\bibitem{must1}O.G. Mustafa, On the uniqueness of flow in a recent tsunami model, Appl{.} Anal{.}
(on-line),\\
\verb+http://dx.doi.org/10.1080/00036811.2011.569499+

\bibitem{podlubny}I. Podlubny, Fractional differential equations, Academic Press, San Diego, 1999

\bibitem{rudin} W. Rudin, Real and complex analysis. Third Edition, McGraw-Hill, New York, 1987

\bibitem{samko} S.G. Samko, A.A. Kilbas, O.I. Marichev, Fractional integrals and derivatives.
Theory and applications, Gordon and Breach, Switzerland, 1993

\end{thebibliography}
\end{document}